\documentclass[10pt,letterpaper]{article}
\usepackage{opex3}
\usepackage{color}
\usepackage[colorlinks=true,urlcolor=blue,linkcolor=black,citecolor=black,breaklinks=true,bookmarks=false]{hyperref}
\usepackage{cite}
\usepackage{amsmath}
\usepackage{siunitx}
\usepackage{braket}
\usepackage{float}
\begin{document}

\title{Ultrafast, high repetition rate, ultraviolet, fiber based laser source: application towards Yb${^+}$ fast quantum-logic}

\author{Mahmood Irtiza Hussain,$^1$ Matthew Joseph Petrasiunas,$^1$ Christopher D. B. Bentley,$^2$ Richard L. Taylor,$^2$ Andr{\'e} R. R. Carvalho,$^{2,3}$ Joseph J. Hope,$^2$ Erik W. Streed,$^{1,4}$ Mirko Lobino,$^{1,5}$ and David Kielpinski$^1$}

\address{$^1$Center for Quantum Dynamics, Griffith University, Nathan, QLD 4111, Australia \\ $^2$Department of Quantum Science, Research School of Physics and Engineering, The Australian National University, Canberra ACT 2601 Australia \\ $^3$Centre for Quantum Computation and Communication Technology, Research School of Physics and Engineering, The Australian National University, Canberra ACT 2601 Australia \\ $^4$Institute for Glycomics, Griffith University, Gold Coast, QLD 4222 Australia\\$^5$Queensland Micro and Nanotechnology Center, Griffith University, Nathan, QLD 4111, Australia} \email{$^*$mahmood.hussain2@griffithuni.edu.au}

\begin{abstract}
Trapped ions are one of the most promising approaches for the realization of a universal quantum computer. Faster quantum logic gates could dramatically improve the performance of trapped-ion quantum computers, and require the development of suitable high repetition rate pulsed lasers. Here we report on a robust frequency upconverted fiber laser based source, able to deliver 2.5 ps ultraviolet (UV) pulses at a stabilized repetition rate of 300.00000 MHz with an average power of 190 mW. The laser wavelength is resonant with the strong transition in Ytterbium (Yb$^+$) at 369.53 nm and its repetition rate can be scaled up using high harmonic mode locking. We show that our source can produce arbitrary pulse patterns using a programmable pulse pattern generator and fast modulating components. Finally, simulations demonstrate that our laser is capable of performing resonant, temperature-insensitive, two-qubit quantum logic gates on trapped Yb$^+$ ions faster than the trap period and with fidelity above 99\%.
\end{abstract}



\section{Introduction}

Trapped ions are envisaged as a future platform for quantum information technologies and for the realization of a full scale universal quantum computer \cite{haffner2008quantum,Monz1068}. In this context an architecture has been proposed \cite{kielpinski2002architecture,home2009complete} for the realization of a scalable, integrated ion trap quantum computer.

However, there are experimental challenges facing its practical realization related to the speed of a two-ion entangling gate which is essential for universal quantum computation. To date, most of the schemes proposed are based on resolving motional side bands of the ion within the trap, which restricts the gate time to be longer than the trap period \cite{cirac1995quantum, poyatos1998quantum}. The fastest entangling gates on trapped ions take several to tens of $\mu$$s$ \cite {leibfried2003experimental,Ballance2014}. Speeding up these gates would represent a major advance toward large-scale quantum computation. 

To overcome this limitation, a gate protocol that uses resonant laser pulses has been proposed \cite{garcia2003speed}. Here the pulses give state-dependent momentum kicks to the ion with one $\pi$ pulse that transfers the population from the ground state to the excited level of the ion, and a second counter-propagating pulse which coherently reverses the process back to the ground state via stimulated emission. Each pulse pair induces a force in the same direction which is proportional to the number of pairs used for the momentum kick. When the timing of the incident pulses is precisely controlled, fast entangling gates can be implemented with minimal motional heating.
Experiments have been demonstrated to realize ultrafast gates with only single-spin operations, with a frequency-tripled multi-watt yttrium vanadate laser producing off-resonant picosecond pulses at a repetition rate of 120~MHz. These pulses are used to drive Raman transitions and to enable quantum logic in a Ytterbium ion (Yb$^+$) \cite{campbell2010ultrafast, mizrahi2013ultrafast}.

\begin{figure*}[!ht]
\centering
\includegraphics[width=\textwidth]{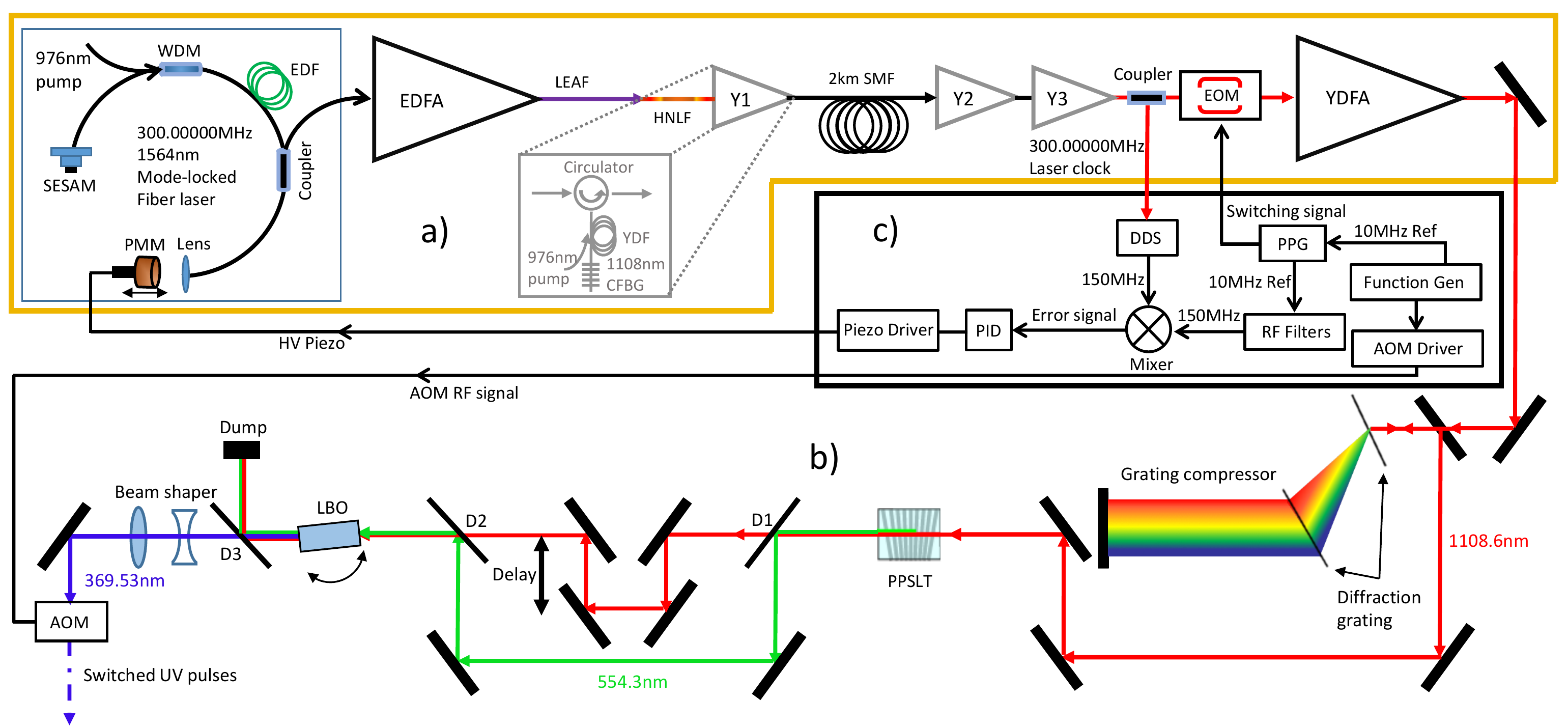}
\caption{Schematic of the laser system. (a) Top yellow panel shows the in fiber portion. Laser pulses generated at 1564nm in the seed oscillator (blue box) are amplified with Erbium doped fiber amplifier (EDFA) and pass through large effective area - highly non linear fibers (LEAF-HNLF) to create an octave-spanning supercontinuum spectrum. Spectral slicing of this spectrum to extract the 1108.6 nm wavelength components takes place in three similar pre-amplification stages (Y1,Y2 and Y3). Ytterbium doped gain fiber, spliced with a chirped fiber Bragg grating (CFBG). Light first enters the circulator which directs it through the gain fiber and then through to the CFBG. Since the gain fiber is before the CFBG, light is amplified twice when passing through each stage. A 2 km long SMF spool between Y1 and Y2 chirps the pulse before high power amplification. A fast switching EOM connected at the output of Y3 is driven by pulse pattern generator (PPG) RF voltage signals. Optical isolators, polarization controllers, fiber polarizer, and EDFA-pre-amplification stage including dispersion compensation fiber are omitted for clarity. (b) The free space up-conversion portion. Chirped pulses from the Ytterbium doped fiber amplifier (YDFA) are compressed with a diffraction grating pair, then the compressed pulses are frequency up-converted to 369.53 nm via PPSLT and LBO crystals. D1, D2, and D3 are dichroic mirrors. (c) Control electronics for pulse pattern generation. Includes a feedback loop to synchronize and stabilize the pulse repetition rate, fast electro-optic-modulator (EOM), and slow acousto-optic-modulator (AOM) drivers for pulse switching signals.}
\label{setup}
\end{figure*}

Here, we demonstrate a fiber laser system designed for ultrafast quantum gate application with Yb$^{+}$ which is tunable to the exact resonance of an atomic transition. The source operates at a repetition rate of 300~MHz which corresponds to the cavity 5th harmonic, and this number can be scaled up by mode locking the cavity to higher order harmonics leading to an increase in the gate speed.

The laser generates 2.5~ps UV pulses, a maximum power of 190~mW and a wavelength that is resonant with the strong, 369.53~nm $S_{1/2}$ - $P_{1/2}$ transition of Yb$^{+}$.
UV pulses are generated starting from an amplified mode-locked telecom band fiber laser emitting 55~fs pulses. Subsequently the pulses propagate in a highly non-linear fiber (HNLF) where a supercontinuum spectrum is generated. A portion of the spectrum around 1108.6~nm is selected from the supercontinuum for the generation of UV power via two frequency upconversion stages in nonlinear crystals. Our architecture allows multiple wavelength channels in the supercontinuum to run in parallel for different applications. For example three stages of frequency doubling of the 1905~nm wavelength could produce a novel source for two photon cooling of hydrogen atoms \cite{PhysRevA.73.063407}. This system architecture offers greater flexibility when compared to commercially available pulsed UV lasers which have a limited range of available repetition rates, wavelengths, and output powers. Except for the harmonic generation stages, laser architecture is fiber based, providing stability and compactness.

\section{Laser source}

\subsection{Mode-locked fiber laser}
The fundamental oscillator in our system is an Er-doped linear fiber laser with a length of $\sim$ 170 cm (see top-left in Fig. \ref{setup}(a)). It operates at 1564~nm wavelength, pumped by a 976~nm laser diode. Fiber lengths are optimized to balance dispersion and non-linearity. Passive mode locking is achieved using a fiber butt coupled, 1.3~mm $\times$ 1.3~mm semiconductor-saturable-absorber mirror (SESAM) with 9\% absorption. The fundamental repetition rate of the cavity is 60~MHz  which can be increased by increasing the pump diode current and altering the cavity birefringence, to produce harmonics in the cavity. Increase in the pump current allows multiple solitons to co-exist in the cavity and cavity birefringence is changed by manual fiber polarization controllers. After optimizations we get a clean and stable (up to many weeks) 5th harmonic and ramp-up the repetition rate to 300~MHz with low timing jitter. To synchronize the laser pulses with the pulse selecting modulators and further minimize the pulse to pulse timing jitter the repetition rate is locked to 300.00000~MHz by introducing a free space piezo-mounted mirror (PMM) on a translation stage at one of the cavity ends.

A repetition rate error signal is produced by RF mixing of an external clock with the laser repetition rate. A direct digital synthesizer (DDS) divides the 300~MHz optical pulse repetition rate by a factor of two to generate a 150~MHz laser repetition rate signal which is then mixed against a 150~MHz reference signal generated by RF-filtering the 15th harmonic from a 10~MHz clock reference. The error signal is fed into a SRS SIM960 proportional-integral-differential (PID) controller which servos the PMM to stabilize the laser's repetition rate. The rest of the laser system inherits this repetition rate.
\subsection{Infrared wavelength selection}
As shown in Fig. \ref{setup}(a) the laser pulses from the fundamental oscillator are amplified to 120mW with an IPG Photonics EAR-IK-C-PM Erbium-doped fiber amplifier (EDFA). The amplified pulses are compressed to 55~fs with 50~cm piece of large effective area fiber (LEAF) and spliced with 14~cm of Sumitomo HNLF. Low soliton number ($<$15) in LEAF and short pulse duration ensure coherence in the supercontinuum and forthcoming stages, which has been detailed previous \cite{kielpinski2009mode} for this system. Pulses passing through the HNLF have sufficient peak intensities to generate an octave-spanning supercontinuum, 1000~nm~$<\lambda<$~2000~nm \cite{kielpinski2009mode}. In the future, the ends of this supercontinuum could be used as a f-2f interferometer to generate a signal useful for stabilization of the seed laser's carrier envelope phase.
\begin{figure}[htbp]
\centering
\includegraphics[width=8cm]{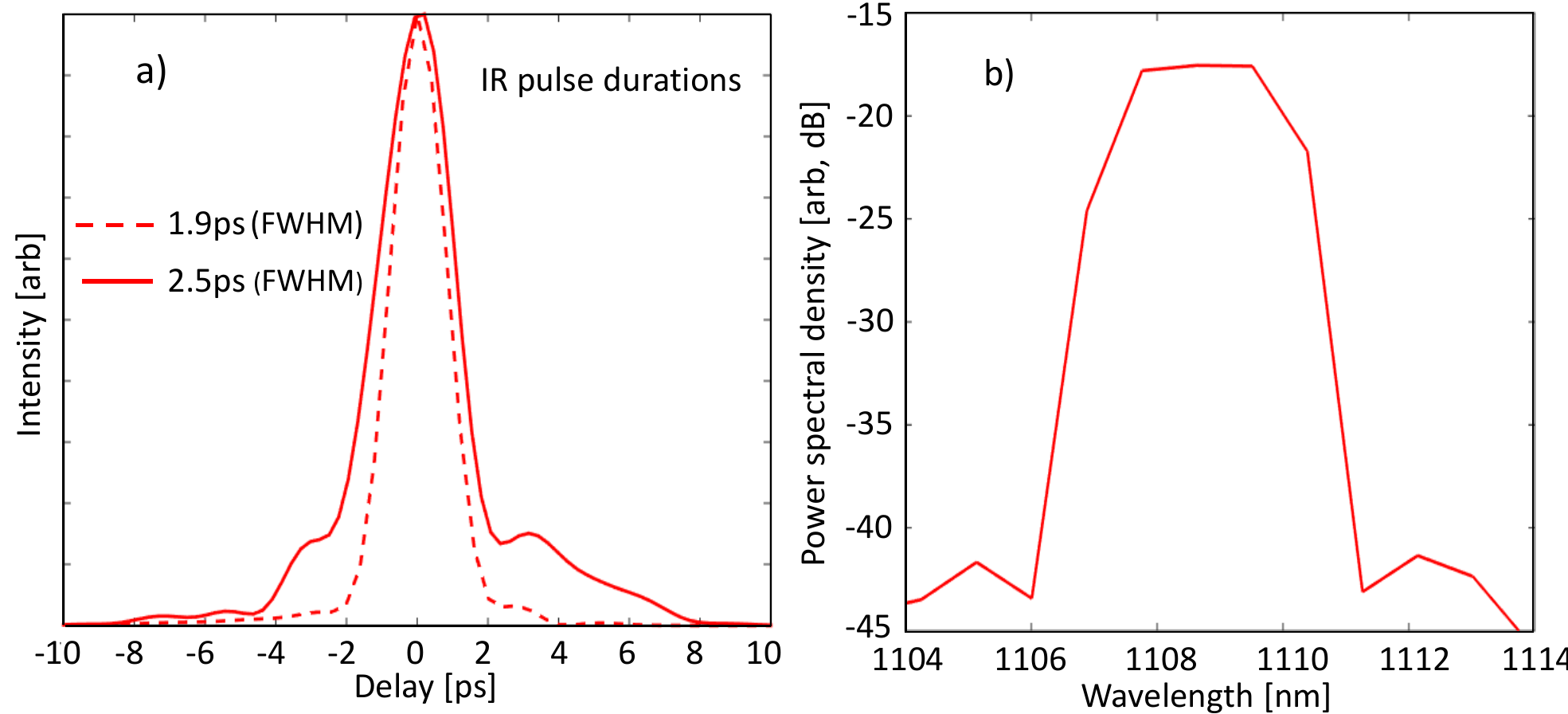}
\caption{(a) Pulse widths of fundamental 1108.6~nm pulses, after grating compressor (dashed line) and after SFG (solid line) used as a reference pulse for cross-correlation measurement. (b) Optical spectrum of supercontinuum after spectral slicing.}
\label{irpulse}
\end{figure} 
Wavelength selection is dependent on the requirements of the desired application. We select 1108.6~nm light from the supercontinuum by first spectral slicing around 1110~nm with a wavelength-division-multiplexer (WDM) and then coarse filtering using a custom-made chirped fiber Bragg grating (CFBG) as shown in Fig. \ref{irpulse}(b). After passing through series of these stages we have a total of $\sim$ 30~mW of average power, sufficient for the high power amplifier. Before amplification, pulses are chirped with 2~km of single mode fiber (SMF), between the first two pre-amplification stages. A Keopsys (F-CYFA-016-01) 5~W Ytterbium-doped fiber amplifier (YDFA) has been used for high power amplification and provides a maximum of 4.7~W average power at output.
\subsection{Ultraviolet pulse generation}

In contrast to the IR amplification stages, for frequency up-conversion we want pulses with the highest intensities in order to obtain a strong nonlinear effects.  The YDFA output is collimated in free space and passes through a pair of LightSmyth LSFSG-1000 transmission gratings, configured with anomalous dispersion (see Fig. \ref{setup}(b)) and an inter-grating distance of about 2.4~m to remove the chirping, which compress the 30~ps pulses to 1.9~ps as indicated in Fig. \ref{irpulse}(a). Frequency resolved optical gating (FROG) was used to measure the pulse length with a MesaPhotonics FROGScan equipped with an Ocean Optics HR4000 spectrometer which covers a wavelength range of $\sim$ 503~nm-577~nm. During dechirping, unwanted amplified spontaneous emission centered at roughly 1070~nm in the YDFA output spectrum is removed by spatial filtering.
The overall compressor efficiency is 75\%, with most losses from multiple passes through the $>$90\% efficient transmission gratings. Hence the maximum average power available for frequency up-conversion of the compressed pulses is 3.5~W.

Up-conversion is a two stage process that starts with second harmonic generation (SHG) of the compressed 1108.6~nm IR laser pulses to green 554.3~nm pulses. The vertically polarized compressed 1108.6~nm pulses are focused to a $\sim$~22~$\mu$m @1/$e^2$ radius spot in a type I, quasi-phase-matched (QPM), periodically-poled-stoichiometric LiTaO$_3$ (PPSLT) crystal, with an estimated  peak intensity  0.38~GW/cm$^2$. We achieved maximum SHG average power of 2~W and a single pass IR to green conversion efficiency of 57\% using an improved version of the apparatus described in \cite{petrasiunas2014picosecond}, with a more powerful YDFA and corresponding dispersion optimization.

\begin{figure}[htbp]
\centering
\includegraphics[width=9cm]{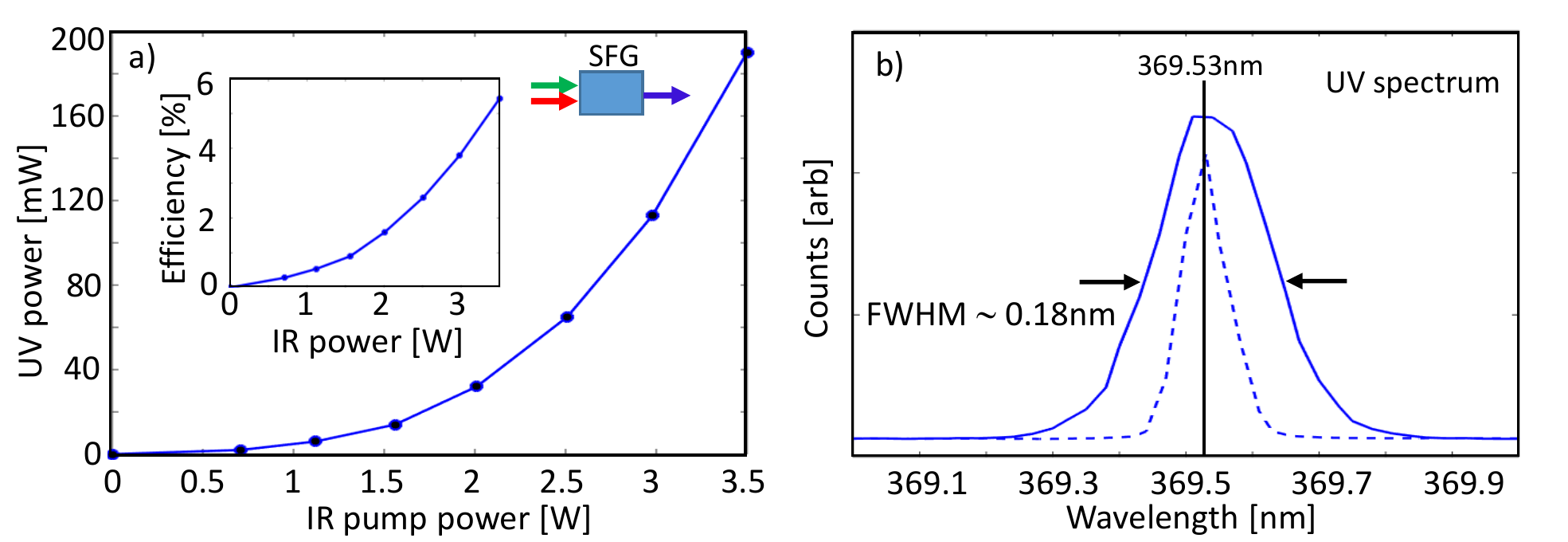}
\caption{(a) SFG power and IR to UV conversion efficiency (inset) versus input fundamental 1108.6~nm pump beam power. (b) UV spectrum of the SFG output centered at 369.53~nm (solid line) and the spectrometer instrument response (dashed line) as measure by a single frequency ($\Delta\lambda<10^{-6}$ nm) UV ECDL locked to an Yb$^+$ atomic reference.}
\label{uvpower}
\end{figure}

After SHG the fundamental and second harmonic beams are separated by a dichroic beamsplitter (Semrock FF665-Di02) with 96\% pump transmission and 98\% reflection for the second harmonic. As shown in Fig. \ref{setup}(b) an adjustable delay is introduced and the beams are then recombined to generate UV pulses via sum frequency generation (SFG). Selection of an appropriate crystal is vital to achieving efficient non-linear conversion of the pump pulses into UV light. This becomes more crucial when dealing with ultrafast pump beams with high repetition rates and therefore limited intensities.

Crystals from the borate family are a promising option for high power UV generation due to their high transparency in the UV spectral region and high optical damage threshold \cite{sasaki2000recent}. Therefore, we use a critically phase matched LiB$_3$O$_5$ (LBO) crystal (10~mm~$\times$~3~mm~$\times$~3~mm, Raicol Crystals, AR coated for all three SFG wavelengths), with a theoretically estimated acceptance bandwidth of $\sim$~1.32~mrad$\cdot$cm \cite{smith2004snlo}, operating in a type I configuration. Second harmonic and residual IR beams are focused to spot sizes of 35~$\mu$m and 40~$\mu$m @ 1/e$^2$ radius respectively inside the LBO. These beams carry maximum peak powers of 1.74~KW and 2.9~KW respectively, corresponding to peak intensities of 35~MW/cm$^2$ and ~57~MW/cm$^2$. In these conditions, we produce a maximum of 190~mW UV power, giving a single pass UV conversion efficiency $\eta = P_{\textrm{370~nm}}/(P_{\textrm{1108~nm}}~\times~P_{\textrm{554~nm}})$ of 8\%W$^{-1}$ in SFG. The lack of saturation roll-off in the UV power graph indicates that the UV conversion efficiency is limited by the input powers, as shown in the Fig. \ref{uvpower}(a). A Semrock Di02-R405-25x36 dichoric beamsplitter and a Semrock FF01-405/150-25 filter (370~nm reflection $>$~90\%) are used to separate the UV light from the IR and green. The SFG-pump beams both have an M$^2$ of $\sim$~1.4, while the resulting UV beam has M$^2>2$ and average circularity $<$~40\%. 

Immediately after isolating the UV beam from the IR and green components, a telescope consisting of cylindrical and plano-convex lenses is placed in the beam path for beam profile correction and collimation, and as a result the UV M$^2$ is improved to 1.5, comparable to the M$^2$ of the input beams. The initial lower beam quality indicates that spatial walk-off in the pump beams is limiting beam overlap and hence conversion efficiency. Spatial walk-off can be reduced to improve SFG efficiency by employing a combination of crystals before LBO \cite{Smith98}. Group velocity mismatch (GVM) could be another factor in degrading UV power due to the long length of the crystal. 

The time bandwidth product (TBWP) of UV pulses is measured to be $\sim$~0.94, as compared to 0.315 for a sech$^2$ shaped pulse. The TBWP is higher than ideal due to uncompensated higher order dispersion in the grating compressor, which could be inferred from chirping in the fundamental IR pulses as shown in Fig. \ref{irpulse}(a). The UV spectrum is centered at 369.53~nm, with a FWHM width of $\sim$~0.18~nm, as measured by an Ocean Optics HR 4000 spectrometer configured for measurement in the UV. We find that the UV spectrum coincides with the spectrum peak of a reference external cavity diode laser (ECDL), locked with the $S_{1/2}$ - $P_{1/2}$ Yb$^+$ transition as shown in Fig. \ref{uvpower}(b). The UV center wavelength can be tuned to the exact resonance by slight variation in the crystal phase matching (within acceptance bandwidth) or by heating/cooling of the wavelength selecting CFBGs. When the laser is fully optimized, UV pulse peak power is stable within 5\% over many hours. The power fluctuations happen on a fast time scale comparable to the laser repetition rate with a pulse timing jitter of <15ps when the repetition rate of the fundamental oscillator is locked and stabilized.

\section{Cross-correlation}

Conventional techniques for characterization of ultrafast pulses rely on the SHG from time delayed copies of pulses. These schemes are limited to IR and visible spectrum. Implementation of such schemes for UV pulses is cumbersome due to unavailability of suitable non-linear crystals and spectrometers. Therefore, in our case UV pulse duration ($\tau_{UV}$) has been estimated by a cross-correlation (CC) technique. After SFG unconverted IR (reference) and UV (to be measured) pulses are separated from the residual green pulses with a Semrock FF01-379/34-25 filter by reflecting 554~nm pulses and transmitting 1108~nm ($>$~70\%) and 370~nm ($>$~90\%) wavelengths.

\begin{figure}[]
\centering
\includegraphics[width=9cm]{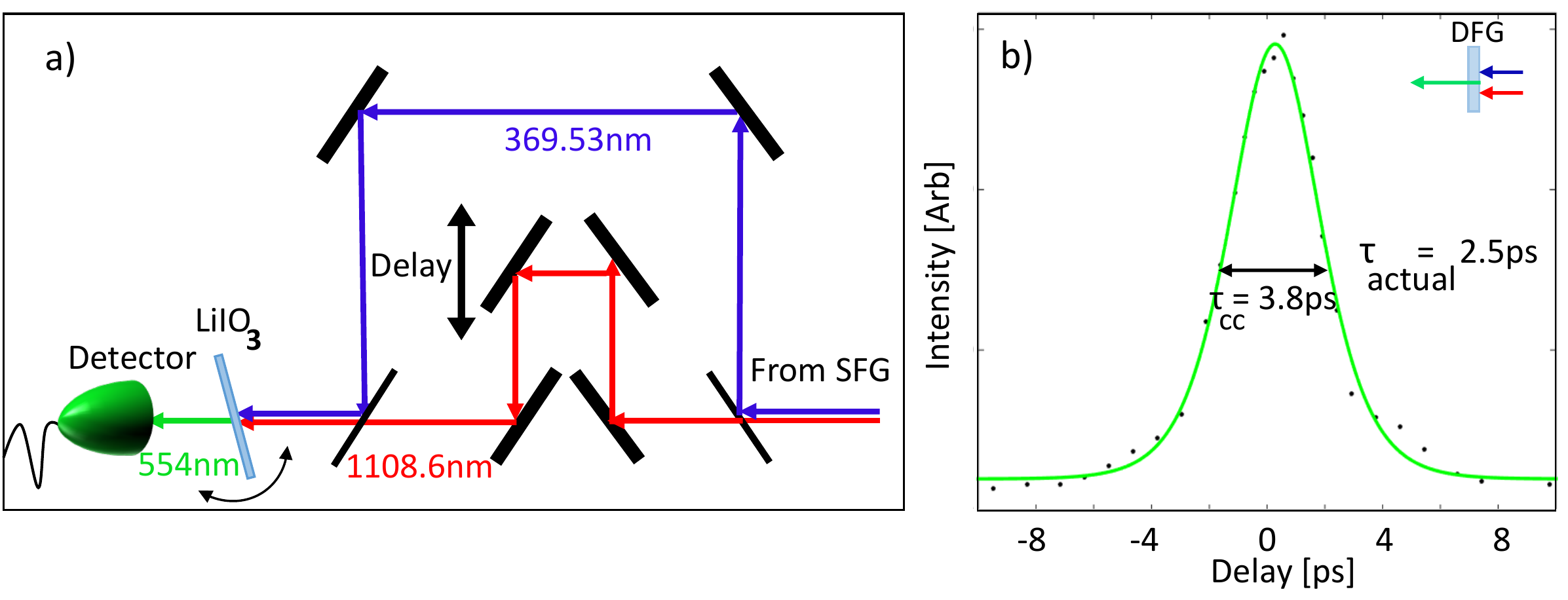}
\caption{(a) Schematic of the setup for cross-correlation measurement, (b) Shows the cross-correlation trace, formed by the sech$^2$ fit to data points, where each data point represents the DFG output power variation against the time delay in the IR pulse.}
\label{cc2}
\end{figure}
Subsequently the IR and UV pulses are separated with a Semrock FF665-DiO2 dichroic beamsplitter with transmission $>$~95\% at 1108~nm and $>$~98\% reflection at 369.5~nm. IR and UV pulses are directed to different paths with a delay line in the IR beam path as shown in Fig. \ref{cc2}(a). Then we combine these pulses again with a dichroic mirror (Semrock FF665-DiO2) and overlap them co-linearly with a beam profiler. Overlapped beams are focused in 1 mm-thick critically phase matched LiIO$_3$ crystal with cut ($\theta$=30.2$^\circ$, $\phi$=0$^\circ$) for type I difference frequency generation (DFG).

As a result of the DFG process a CC signal at 554~nm is produced. Delay translation in small steps provides us with different signal strengths that form the CC envelope, with a FWHM of 3.8 $\pm 0.1$~ps as shown in Fig.~\ref{cc2}(b). To extract the unknown pulse duration from the CC envelope duration, their deconvolution is needed. First we measured the FWHM duration of the reference pulses ($\tau_{IR}$ = 2.5~ps) using FROG and from the CC FWHM duration ($\tau_{CC}$) of 3.8~ps we estimated a $\sim$~2.5~ps duration of the UV pulses. For negligible GVM, we can calculate ${\tau_{UV}} = ({\tau_{CC}^{p} - \tau_{IR}^{p}})^{p^{-1}}$ to be $\sim$~2.5~ps, where $p^{-1}$ $\sim$~0.6 for sech$^2$ pulses \cite{baronavski1993analysis}. Also, the actual-CC and reference pulse durations are roughly the same, which gives us a straight forward estimate of UV pulse duration at $\sim$~2.5~ps.

IR pulse broadening can be attributed to the dispersion induced chirping after SHG and SFG, as evident from pedestal formation as shown in the inset of Fig.~\ref{irpulse}(a).

\section{Pulse switching}
Generating arbitrary UV pulse patterns by switching on and off individual pulses is one of the most critical steps for the implementation of fast quantum logic gates. The high power, ultrafast duration and high repetition rate of the pulses make it extremely difficult to achieve fast switching in a single stage with currently available technology.

\begin{figure}[!]
\centering
\includegraphics[width=9cm]{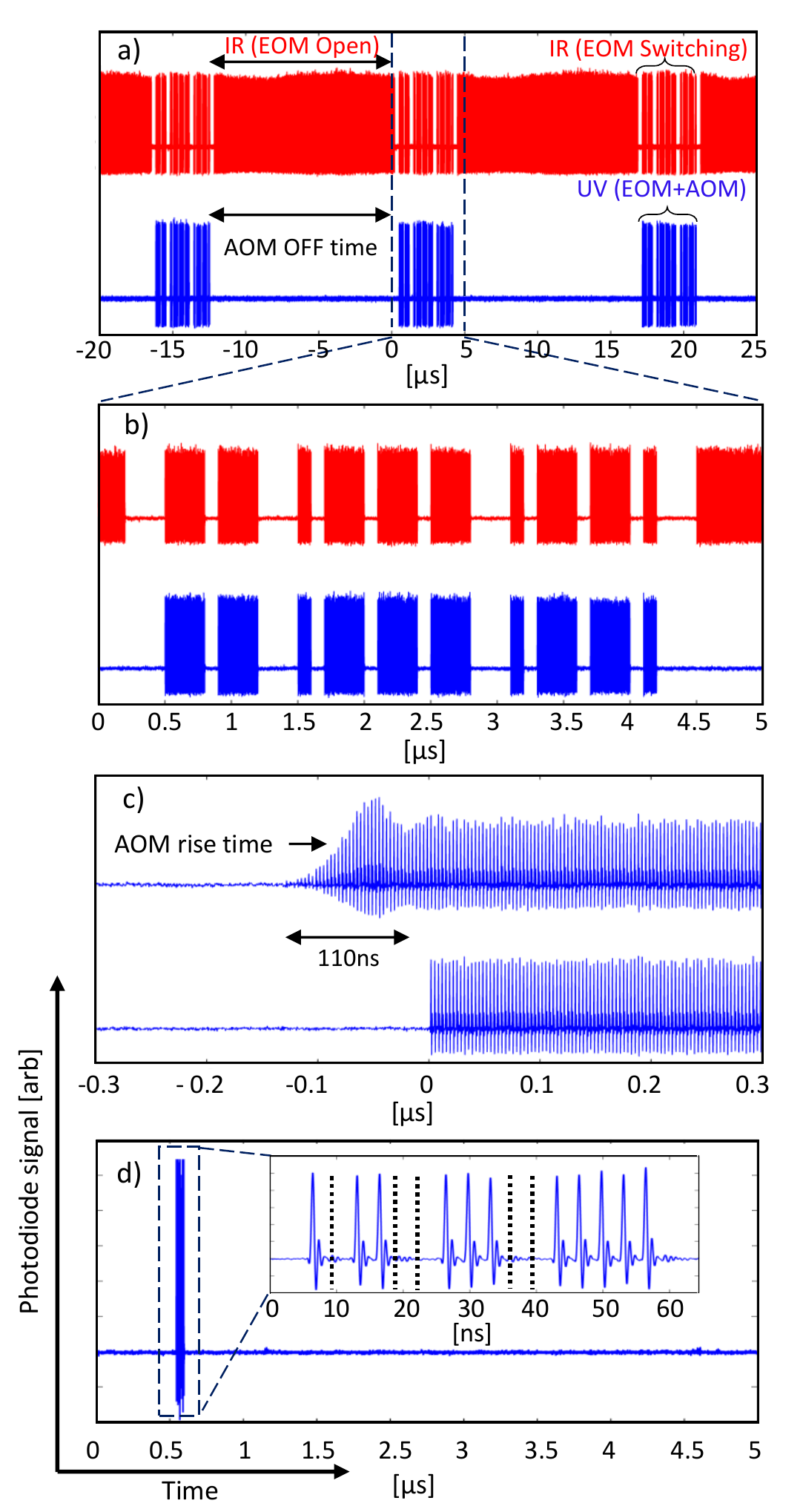}
\caption{(a) Many-pulse pattern: the red (top) pulse train represents the EOM-switched 1250 IR pulses and the bottom (blue) represents the UV pulse pattern formed after synchronizing the EOM and AOM switching. (b) Shows the enlarged image of the many-pulse switching pattern. (c) the UV pulse train (top) shows AOM switching of UV pulses without EOM switching and the bottom shows EOM switching to balance the rise time of the AOM to ensure a clean pulse switching pattern. (d) Few-pulse pattern: the EOM- and AOM-switched UV pulse patterns, showing the ability to individually switch UV pulses to write patterns within the main pattern. Black dashed lines indicate OFF pulses. The ringing of the photodiode response shown above is an artifact due to the limited bandwidth of the detector with respect to the pulse duration.}
\label{switch}
\end{figure}

To overcome this problem we have instead implemented a two-stage solution based on fast low-power IR switching and slower high power UV switching to generate almost arbitrary patterns. For fast IR switching we use an EOspace polarization-maintaining  lithium niobate electro-optic modulator (EOM) between the last pre-amplification stage and the YDFA. The EOM has insertion losses $<$~4~dB and requires a RF driving voltage of V$_{\pi}$~$<$~3.5~V. The pulse sequences are generated by digitally modulating the EOM with amplified and DC biased pulses from a 12.5~GHz Picosecond Pulse Labs Model 12050 pulse pattern generator (PPG). The PPG operating in the data configuration mode writes out a pattern of digital pulses (1/0), which correspond to the output RF voltage states for EOM (ON/OFF). Fig. \ref{switch}(a) (red) shows a pulse pattern generated by the PPG. This pulse modulation propagates through all subsequent stages in the laser system until the UV acousto-optic modulator (AOM). We measured the spectral and temporal properties of the switched pulses using FROG at different points during up-conversion and after DFG, and find that the pulse properties remain the same.

For pulse sequences with relatively long off periods or a low duty cycle in output the IR EOM is not a complete solution. The IR EOM can only be shut for up to~10\% of the total time since the YDFA stage requires a minimum average input power to prevent the amplifier from being damaged. To implement arbitrary pulse switching we also modulate the UV pulses directly with an AA Opto-Electronics Model MQ110-A3-UV  AOM (see Fig. \ref{switch}(a), blue). The modulator is driven with an RF power of $<$~4~W at 110~MHz, and has a 1st order diffraction efficiency of~85\%.

Synchronization of the fast and slow modulators with respect to the pulse train is essential to generate complex UV pulse patterns. The AOM function generator produces a 10~MHz clock reference fed in to the PPG. The PPG provides the reference signal for the repetition rate stabilization which aligns single pulses and individual switching windows with a fixed delay.  
As a result, a fast-switched UV pulse pattern has been generated with long time delays as shown in Fig.~\ref{switch}. Time-domain pulse patterns are recorded by a DSO8104A Agilent fast oscilloscope with a GHz resolution bandwidth. We intentionally introduce a dead time of 45 pulses (150~ns) at the start and end of the fast switching program to balance the rise and fall time in the AOM (110 ns) as shown in Fig.~\ref{switch}(c). Besides many-pulse UV patterns as discussed so far, few-pulse UV patterns with OFF time $>$~4~$\mu$s are also produced as shown in Fig.~\ref{switch}(d). Using the few-pulse procedure it is also possible to make short arbitrary pulse patterns followed by many $\mu$s OFF time, such as required for implementing quantum gates, which illustrates the flexibility of the switching system in generating a variety of pulse patterns.

\section{Fast gate implementation analysis}
In order to assess the application of this laser system to ultrafast quantum logic \cite{garcia2003speed,Bentley2013,Bentley2015} in Yb$^+$, we calculate the fidelity for a fast gate performed with the given laser parameters, following the techniques presented in \cite{Bentley2016}. Table \ref{tab:fastgate} shows the results given the laser repetition rate of 300 MHz and Gaussian intensity fluctuations of 2.67\% standard deviation around the average intensity to mimic the experimentally measured fluctuations. The Fast Robust Antisymmetric Gate (FRAG) \cite{Bentley2013,Bentley2015} was found to be optimal for this setup, and two gates are presented: a 30 pulse-pair gate with a duration of 2.8 $\mu$s, and 150 pulse-pair gate with a duration of 0.55 trap periods or 693 ns. The 150 pulse-pair gate is the fastest FRAG scheme possible with a 300 MHz repetition rate. The mean and standard deviation in fidelity are given for each gate and different initial number states $n$ in Table \ref{tab:fastgate}, where the fidelity distribution arises from Gaussian fluctuations in the laser intensity and thus pulse area. The small pulse time jitter of the laser does not affect the gate fidelity.

\begin{table}[!htb]
\caption{Expected mean fidelity and standard deviation in fidelity for two examples of fast gates that can be performed with this laser, incorporating laser intensity fluctuations.}\label{tab:fastgate}
\centering
\begin{tabular}{|l|l|l|}
\hline
\multicolumn{3}{|l|}{30 pulse pairs, $T_G$ = 2.8~$\mu$s}\\
\hline
Motional state $n$ & Fidelity & Standard deviation\\
\hline
0 & 0.9998 & 0.0001\\
1&0.9996&0.0002\\
2&0.9993&0.0003\\
3&0.9991&0.0003\\
4&0.9989&0.0004\\
\hline
\multicolumn{3}{|l|}{150 pulse pairs, $T_G$ = 693~ns}\\
\hline
Motional state $n$ & Fidelity & Standard deviation\\
\hline
0&0.9987&0.0004\\
1&0.9977&0.0007\\
2&0.9964&0.0010\\
3&0.9951&0.0015\\
4&0.9940&0.0019\\
\hline
\end{tabular}

\end{table}
Fidelity calculations were performed using a linear Paul trap containing two Yb$^+$ ions, with trap frequency $\nu=\frac{5}{2\pi}$~MHz. A representative internal state, $|\psi_0\rangle_i = \frac{1}{\sqrt{2}} (|00\rangle + |01\rangle )$, was chosen to calculate the fidelity and capture the relative phase dynamics of the gate.
The initial motional state of the trap was taken to be a tensor product of Fock states with equal $n$ for the centre-of-mass and stretch modes,
\begin{equation}
|\psi_0\rangle_m=|n\rangle_\textrm{c.m.}\otimes|n\rangle_\textrm{st.},
\end{equation}
which simplifies computational requirements for fidelity calculation \cite{Bentley2016}.
The consistent fidelities for each initial number state $n$ provide a conservative approximation for an initial low-temperature thermal state, due to population in such a state being dominated by low-$n$ excitations in the number basis.

The counter-propagating $\pi$-pulse pairs comprising the fast gate provide momentum kicks at times determined by the FRAG scheme. Pulse switching will provide the required freedoms in pulse timings.
By applying a $\pi$ phase shift to the second pulse in each pair \cite{Kregar2014}, the Bloch sphere rotation for each ion occurs in the opposite direction to the rotation from the first pulse.  Area fluctuations in each pair of pulses are correlated, as the pulse pairs are constructed from a single laser pulse using a beam-splitter; the $\pi$ phase shift thus cancels unwanted population transfer.  The mean pulse area is assumed to perform an ideal $\pi$ pulse on the ions, however systematic errors are similarly canceled by the phase shift.

The simulation results show that this laser system is already capable of performing entangling gates, with fidelity above 99.9\%, that are faster than demonstrated gates in a trapped ion system.  A gate twice as fast as the trap period can be performed with fidelity above 99\%.  Fidelity is dependent upon the trap's initial motional state; higher temperatures exacerbate infidelity from inexact momentum transfers.  To provide fidelity improvements above 99\% or to extend the scheme to higher temperatures, composite pulse schemes can be used to construct $\pi$ pulses with increased robustness to intensity fluctuations \cite{Ivanov2011}.

\section{Conclusions and outlook}
In conclusion, we have demonstrated a pulsed fiber laser with scalable repetition rate, capable of generating picosecond UV pulses at 300~MHz. Our laser is ideal for high-fidelity fast gates in ion traps, where the pulse repetition rate must be far greater than the trap frequency. The fiber architecture provides robustness and compactness to the laser system, which can be further increased by replacing the free-space pulse compressor with a hollow-core fiber. The octave-spanning supercontinuum contains a range of wavelengths, which enables ample choice of fundamental wavelengths. We choose 1108.6~nm and frequency upconversion to 369.53~nm via SHG and SFG, for resonance with the $S_{1/2}$ - $P_{1/2}$ energy levels of the Yb$^+$ ion. The maximum single-pass SFG efficiency is 8\%~W$^{-1}$ and the overall IR-UV conversion efficiency is 5.4\%, which compares quite favorably with the previously achieved efficiencies with various peak pump intensities \cite{li2014high,chaitanya2015stable,chaitanya2015high}. Besides many potential applications, this laser has specifically been modified to operate as a fast quantum-logic gate source. For this modification we have developed a fast UV pulse switching technique to produce arbitrary pulse patterns in time. The UV pulse duration of 2.5~ps has been estimated by cross-correlating UV pulses with the residual IR pulses via DFG after SFG. 

Scalability in the repetition rate without significantly changing the pulse energy is one of the distinguishable features of our laser system. With modification to our current set-up it is possible to push the repetition to the GHz range and keep the pulse energy the same. The laser design would not necessarily have to change, but to keep the same pulse energy would require commensurate scaling of the average power. Recent progress \cite{petersen2015large} in fiber amplifier technology makes this prospective feature of laser system feasible.

Simulations of gate fidelities show that the laser is capable of performing high-fidelity fast gates with gate times faster than previous implementations, and even below the trap period. Higher repetition rates would provide further improvements in the gate time: simulations for a 1 GHz repetition rate with our laser's parameters provide gate times as fast as 0.36 trap periods (447 ns) with fidelity above 99\%.
\section*{Funding information}
This work was supported by the Australian Research Council through the following programs: Centre of Excellence for Quantum Computation and Communication Technology CE110001027, Discovery Project DP130101613, Future Fellowships FT120100291 (JJH), FT110100513 (DK), and FT130100472 (EWS), and Discovery Early Career Research Award (DECRA) DE130100304 (ML).
 
\section*{Acknowledgments}

The authors thank John Canning and Michael Stevenson for providing chirped fiber Bragg gratings. We also thank Rainer Blatt for useful discussion. 


\begin{thebibliography}{99}

\bibitem{haffner2008quantum} H.~H{\"a}ffner, C.~F.~Roos, and R.~Blatt, ``Quantum computing with trapped ions,'' Phys. Rep. {\bf 469}(4), 155--203 (2008).
\bibitem{Monz1068} T.~Monz, D.~Nigg, E.~A. Martinez, M.~F.~Brandl, P.~Schindler, R.~Rines, S.~X.~Wang, I.~L.~Chuang, R.~Blatt, ``Realization of a scalable Shor algorithm,'' Science {\bf 351}(6277), 1068--1070 (2016).
\bibitem{kielpinski2002architecture} D.~Kielpinski, C.~Monroe, and D.~J.~Wineland, ``Architecture for a large-scale ion-trap quantum computer,'' Nature {\bf 417}(6890), 709--711 (2002).
\bibitem{home2009complete} J. P. Home, D. Hanneke, J. D. Jost, J. M. Amini, D. Leibfried,
and D. J. Wineland, ``Complete methods set for scalable ion trap quantum information processing,'' Science {\bf 325}(5945), 1227--1230 (2009).

\bibitem{cirac1995quantum} J. I. Cirac and P. Zoller, ``Quantum computations with cold trapped ions,'' Phys. Rev. Lett. {\bf 74}(20), 4091 (1995).

\bibitem{poyatos1998quantum} J. F. Poyatos, J. I. Cirac, and P. Zoller, ``Quantum gates with “hot” trapped ions,'' Phys. Rev. Lett. {\bf 81}(6), 1322 (1998).

\bibitem{leibfried2003experimental} D. Leibfried, B. DeMarco, V. Meyer, D. Lucas, M. Barrett,
J. Britton,W. Itano, B. Jelenkovi´c, C. Langer, T. Rosenband, D. J. Wineland, ``Experimental demonstration of a robust, high-fidelity
geometric two ion-qubit phase gate,'' Nature {\bf 422}(6930), 412--415 (2003).

\bibitem{Ballance2014} C. Ballance, T. Harty, N. Linke, and D. Lucas, ``High-fidelity
two-qubit quantum logic gates using trapped calcium-43 ions,'' arXiv preprint arXiv:1406.5473 (2014).

\bibitem{garcia2003speed} J. J. Garcia-Ripoll, P. Zoller, and J. I. Cirac, ``Speed optimized
two-qubit gates with laser coherent control techniques for ion trap quantum computing,'' Phys. Rev. Lett. {\bf 91}(15), 157901 (2003).

\bibitem{campbell2010ultrafast} W. Campbell, J. Mizrahi, Q. Quraishi, C. Senko, D. Hayes,
D. Hucul, D. Matsukevich, P. Maunz, and C. Monroe, ``Ultrafast gates for single atomic qubits,'' Phys. Rev. Lett. {\bf 105}(9), 090502 (2010).

\bibitem{mizrahi2013ultrafast} J. Mizrahi, C. Senko, B. Neyenhuis, K. Johnson, W. Campbell, C. Conover, and C. Monroe, ``Ultrafast spin-motion entanglement and interferometry with a single atom,'' Phys. Rev. Lett. {\bf 110}(20), 203001 (2013).

\bibitem{PhysRevA.73.063407} D. Kielpinski, “Laser cooling of atoms and molecules with ultrafast pulses,'' Phys. Rev. A {\bf 73}(6), 063407 (2006).
\bibitem{kielpinski2009mode} D. Kielpinski, M. Pullen, J. Canning, M. Stevenson, P. Westbrook, and K. Feder, ``Mode-locked picosecond pulse generation
from an octave-spanning supercontinuum,'' Opt. Express {\bf 17}(23), 20833--20839 (2009).

\bibitem{petrasiunas2014picosecond} M. Petrasiunas, M. I. Hussain, J. Canning, M. Stevenson, and D. Kielpinski, ``Picosecond 554 nm yellow-green fiber
laser source with average power over 1 W,'' Opt. Express {\bf 22}(15), 17716--17722 (2014).

\bibitem{sasaki2000recent} T. Sasaki, Y. Mori, M. Yoshimura, Y. K. Yap, and T. Kamimura, ``Recent development of nonlinear optical
borate crystals: key materials for generation of visible and UV light,'' Materials Science and Engineering: R: Reports {\bf 30}(1), 1--54 (2000).

\bibitem{smith2004snlo} A. V. Smith, ``SNLO nonlinear optics code,'' Sandia National Laboratories, Albuquerque, NM {\bf 87185}, 1423 (2004).

\bibitem{Smith98} A. V. Smith, D. J. Armstrong, and W. J. Alford, ``Increased acceptance bandwidths in optical frequency conversion by use of multiple walk-off-compensating nonlinear crystals,'' J. Opt. Soc. Am. B {\bf 15}(1), 122--141 (1998).

\bibitem{baronavski1993analysis} A. P. Baronavski, H. D. Ladouceur, and J. K. Shaw, ``Analysis of cross correlation, phase velocity mismatch and group velocity mismatches in sum-frequency generation,'' IEEE J. Quantum Electron. {\bf 29}(2), 580--589 (1993).


\bibitem{Bentley2013} C. D. B. Bentley, A. R. R. Carvalho, D. Kielpinski, and J. J. Hope, ``Fast gates for ion traps by splitting laser pulses,''
New J. Phys. {\bf 15}(4), 043006 (2013).

\bibitem{Bentley2015} C. D. B. Bentley, A. R. R. Carvalho, and J. J. Hope, ``Trapped ion scaling with pulsed fast gates,'' New J. Phys. {\bf 17}(10), 103025 (2015).

\bibitem{Bentley2016} C. D. B. Bentley, R. L. Taylor, A. R. R. Carvalho, and J. J. Hope, ``Stability thresholds and calculation techniques for fast entangling gates on trapped ions,'' Phys. Rev. A {\bf 93}(4), 042342 (2016).

\bibitem{Kregar2014} G. Kregar, N. {\v{S}}anti{\'c}, D. Aumiler, H. Buljan, and T. Ban, ``Frequency-comb-induced radiative force on cold rubidium atoms,'' Phys. Rev. A {\bf 89}(5), 053421 (2014).

\bibitem{Ivanov2011} S. S. Ivanov and N. V. Vitanov, ``High-fidelity local addressing of trapped ions and atoms by composite sequences of
laser pulses.'' Opt. Lett. {\bf 36}(7), 1275--1277 (2011).

\bibitem{li2014high} K. Li, L. Zhang, D. Xu, G. Zhang, H. Yu, Y. Wang, F. Shan, L. Wang, C. Yan, Y. Wu, X. Lin, J. Yao, ``High-power picosecond 355 nm laser based on $\uppercase{L}a_2 \uppercase{C}a\uppercase{B}_{10}\uppercase{O}_{19}$ crystal,'' Opt. Lett. {\bf 39}(11), 3305--3307 (2014).

\bibitem{chaitanya2015stable} S.~C.~Kumar, E.S. Bautista, and M. Ebrahim-Zadeh, ``Stable, high-power, Yb-fiber-based, picosecond ultraviolet generation at 355 nm using $\uppercase{B}i\uppercase{B}_3\uppercase{O}_6$,'' Opt. Lett. {\bf 40}(3), 403--406 (2015).

\bibitem{chaitanya2015high} N. A. Chaitanya, A. Aadhi, M. V. Jabir, and G. k. Samanta, ``Highpower, high-repetition-rate, yb-fiber laser based femtosecond source at 355 nm,'' Opt. Lett. {\bf 40}(18), 4269--4272 (2015).

\bibitem{petersen2015large} S. R. Petersen, M. Chen, A. Shirakawa, C. B. Olausson, T. T. Alkeskjold, and J. Laegsgaard, ``Large-mode-area hybrid
photonic crystal fiber amplifier at 1178 nm,'' Opt. Lett. {\bf 40}(8), 1741--1744 (2015).

\end{thebibliography}
\end{document}